\documentclass{emulateapj}
\usepackage[ansinew]{inputenc}
\usepackage{hyperref}
\usepackage{graphicx}
\usepackage[dvipsnames]{xcolor}
\usepackage{lineno}
\usepackage{rotating}
\usepackage{ulem}

\definecolor{gray100}{RGB}{3,68,84}
\definecolor{gray50}{RGB}{11,120,147}
\definecolor{gray35}{RGB}{147,228,247}
\definecolor{gray0}{RGB}{223,248,254}

\def\metersecond{$\rm m\,s^{-1}$}

\def\micrometer{$\rm \mu m$}

\begin{document}

\title{Free Collisions in a Microgravity Many-Particle Experiment. III. \\The collision behavior of sub-millimeter-sized dust aggregates}

\author{Stefan Kothe$^1$, J{\"u}rgen Blum$^1$, René Weidling$^1$, and Carsten G{\"u}ttler$^{2,1}$}

\affil{1 Institut f{\"u}r Geophysik und extraterrestrische Physik, Technische Universit{\"a}t zu Braunschweig,\\Mendelssohnstr. 3, D-38106 Braunschweig, Germany\\
2 Department of Earth and Planetary Sciences, Kobe University 1-1 Rokkodai, Nada Kobe 657-8501,Japan}
\email{s.kothe@tu-bs.de}

\begin{abstract}

We conducted micro-gravity experiments to study the outcome of collisions between sub-mm-sized dust agglomerates consisting of $\mu$m-sized $\mathrm{SiO_2}$ monomer grains at velocities of several $\mathrm{cm\,s^{-1}}$. Prior to the experiments, we used X-ray computer tomography (nano-CT) imaging to study the internal structure of these dust agglomerates and found no rim compaction so that their collision behavior is not governed by preparation-caused artefacts. We found that collisions between these dust aggregates can lead either to sticking or to bouncing, depending mostly on the impact velocity. While previous collision models derived the transition between both regimes from contact physics, we used the available empirical data from these and earlier experiments to derive a power law relation between dust-aggregate mass and impact velocity for the threshold between the two collision outcomes. In agreement with earlier experiments, we show that the transition between both regimes is not sharp, but follows a shallower power law than predicted by previous models \citep{GuettlerEtal:2010}. Furthermore, we find that sticking between dust aggregates can lead to the formation of larger structures. Collisions between aggregates-of-aggregates can lead to growth at higher velocities than homogeneous dust agglomerates.

\end{abstract}


\section{Introduction}\label{section_Introduction}

Within the past years, close collaborations between experimental and theoretical physicists enriched our understanding of the processes which finally lead to the formation of planetesimals. From the experimenters' side, many experiments on the collisions behavior (see the experiment reviews by \citet{BlumWurm:2008} and \citet{GuettlerEtal:2010}) and mechanical properties \citep{GuettlerEtal:2009} of porous dust agglomerates were conducted. Laboratory experiments have shown that collisions can lead to various outcomes. Dust agglomerates will stick, due to the inelasticity of the collisions and the attractive van-der-Waals force, as long as their collision velocity does not exceed the bouncing threshold velocity. Sticking can occur either at the first contact at very low velocities or after some plastic deformation in the contact region. \citet{GuettlerEtal:2010} developed an experiment-based dust-aggregate collision model, which predicts the outcome of collisions between dust agglomerates as a function of the masses, porosities, and collision velocity of the dust aggregates.
Based on this model, \citet{ZsomEtal:2010a} simulated the growth of dust aggregates in various protoplanetary-disk (PPD) models, using a Monte-Carlo code. The simulations by \citet{ZsomEtal:2010a} showed that, depending on the PPD model, rather compact dust aggregates of sizes up to one centimeter can grow at 1 AU. However, the growth always stops due to the occurrence of bouncing in dust-aggregate collisions, which reduces the porosity and does not lead to a further increase in aggregate mass. This has been referred to as the bouncing barrier. More recent work by \citet{WindmarkEtal:2012a,WindmarkEtal:2012b} showed that the bouncing barrier can be overcome either by the introduction of a few bigger particles or by taking a Maxwellian distribution of collision velocities into account. Both processes lead to the onset of a fragmentation-induced mass transfer from smaller to larger bodies.
\par
However, it has to be noted that there were only a few experiments which studied collisions between dust aggregates close to or at the threshold between sticking and bouncing so that all previous outcomes are somehow model dependent and not empirically proven. Recently, \citet{WeidlingEtal:2012} (in the following referred to as Paper I) studied collisions between mm-sized dust agglomerates. They found that the transition between sticking and bouncing was not sharp, but followed a wide transition from perfect sticking to perfect bouncing, spanning a range of two orders of magnitude in velocity. As the impact velocities in the experiments by \citet{WeidlingEtal:2012} were not sufficiently low to study the whole transition range, they modified a collision model for plastic/elastic spheres \citep{ThorntonNing:1998} for dust aggregates, taking into account the porous surfaces of the colliding bodies. They confirmed the original threshold between sticking and bouncing introduced by \citet{GuettlerEtal:2010} as the line where half of the dust aggregates stick.
\par
To gain a deeper insight into the physics of dust-aggregate collisions, numerical simulations have made increasing contributions. Internal processes, like restructuring and the formation of cracks in a collision, are not easily observable in laboratory experiments but are readily available in numerical simulations. They also allow for the study of collisions of dust agglomerates with internal structures (e.g., very high porosities, well-defined inhomogenities) that are hardly feasible in experiments. However, one major discordance between experiments and simulations has not been clarified so far: in most cases simulations are not able to achieve bouncing in the way it was observed in several experiments \citep{HeisselmannEtal:2007, WeidlingEtal:2012,BlumMuench:1993}. In these cases, collisions between dust aggregates lead to the rebound of the particles, without any visible mass loss, mass transfer, or deformation of the dust aggregates.
\par
\citet{GeretshauserEtal:2012} compared the collision outcomes observed and categorized by \citet{GuettlerEtal:2010} with the results from their smoothed particle hydrodynamics (SPH) model, which had been calibrated with laboratory experiments \citep{GuettlerEtal:2009, GeretshauserEtal:2010}. \citet{GeretshauserEtal:2012} were able to observe bouncing collisions in their simulations. They collided spherical dust agglomerates with 10 cm and 6 cm diameter and an initial volume filling factor of $\phi=0.35$ (i.e., with a porosity of 65\%) at different velocities. The simulations showed that the particles bounce at velocities of less than 1 \metersecond. The colliding dust aggregates are visibly compacted to $\phi = 0.45$ within the contact volume. At higher impact velocities, sufficient energy is dissipated in restructuring so that the contact area between the two spheres becomes large enough to lead to sticking. This is in contrast to the experimental observations collected in \citet{GuettlerEtal:2010} where sticking only occurs at (much) lower velocities. Sticking in collisions between dust aggregates at high velocities was experimentally only observed if one of the dust aggregates was much smaller (projectile) than the other one (target). In these cases, the deep penetration of the projectile into the target agglomerate leads to the capturing of the projectile \citep{LangkowskiEtal:2008}. Experiments with cm-sized or larger dust agglomerates with various porosities have shown no sign of sticking at or around 1 \metersecond\ \citep{BeitzMeisneretal:2011,SchraeplerEtal:2012}. In contrast, all laboratory experiments have confirmed the transition between bouncing and fragmentation at a velocity of about 1 \metersecond\ velocity \citep{GuettlerEtal:2010,BeitzMeisneretal:2011,SchraeplerEtal:2012}.
\par
\begin{figure}[ptb]
    \includegraphics[width=\columnwidth]{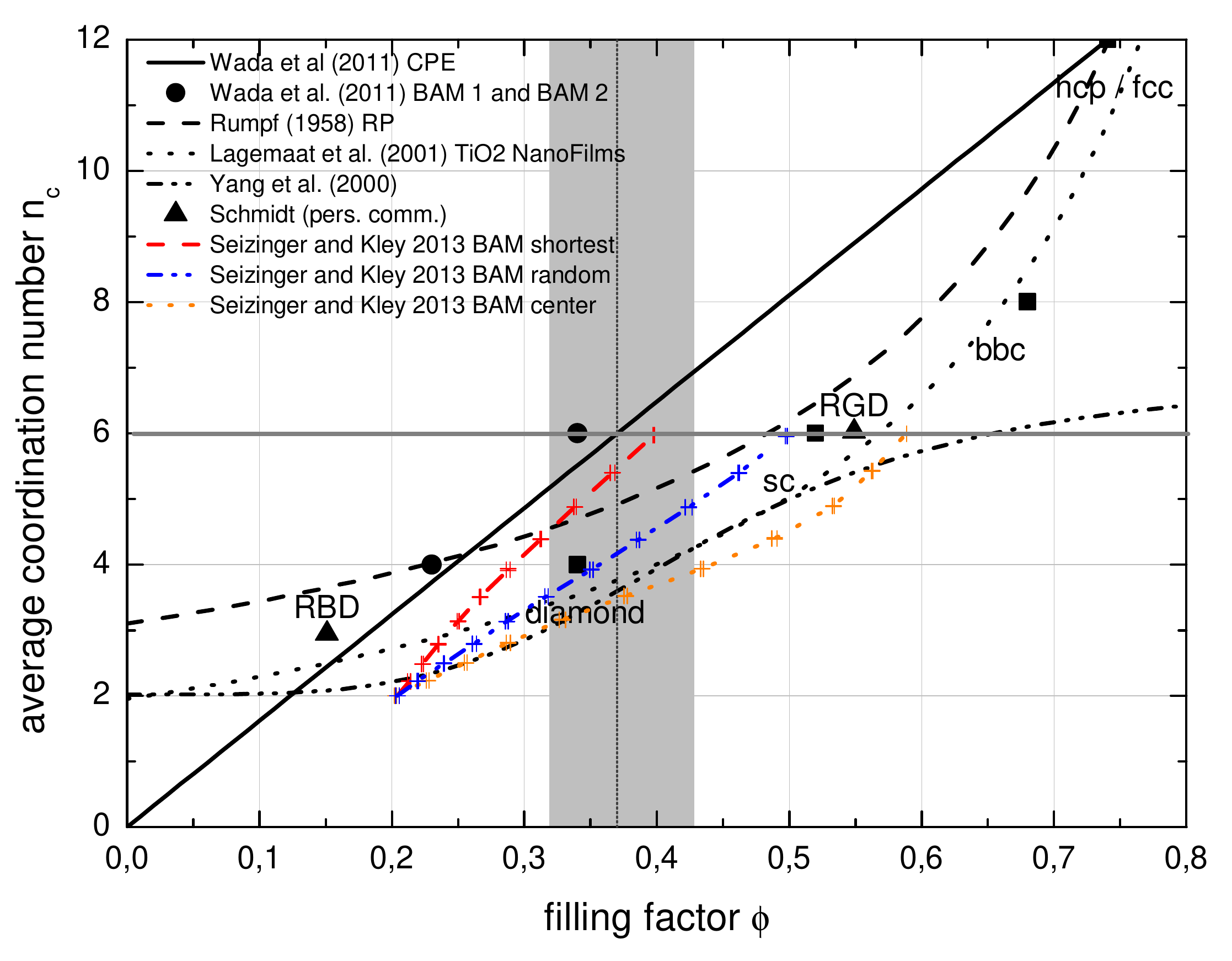}
    \caption{Comparison of different numerical models describing the relation between coordination number and volume filling factor of packed particles. The triangles represent the simulation from Schmidt and Blum (pers. comm.) for random ballistic deposition (RBD) and random gravitational deposition (RGD). The dotted vertical line represents the average volume filling factor of our dust agglomerates and the shaded range denotes the $\pm 1\sigma$ range. The squares represent the coordination number and filling factor of different lattices (diamond, simple (SC), body-centered (bcc) and face-centered (fcc) cubic, resp. hexagonal closed packing (hcp)). }
    \label{fig.coordination_number}
\end{figure}

Rebound in central collisions has also been studied in the numerical simulations by \citet{WadaEtal:2011}. They used a soft-sphere molecular-dynamics model to simulate collisions between dust agglomerates consisting of $ \sim 4000$ monomer grains with various filling factors and collision velocities. The monomers were spherical $\mathrm{SiO_2}$ particles with a diameter of 0.1 $\rm \mu m$. The porosity of dust agglomerates can be described by the coordination number, which is the average number of (touching) next neighbors of a monomer grain. For the case of a close-packed hexagonal structure, from which a fraction $f_\mathrm{ex}$ of particles are randomly withdrawn, \citet{WadaEtal:2011} give the following dependence of the volume filling factor on the coordination number $\mathrm{n_c}$:
\begin{equation}
\label{Eq.nc}
    n_\mathrm{c} =12 \cdot (1-f_{\mathrm{ex}})
    = 12 \cdot \frac{\phi}{\phi_{\mathrm{cp}}} ,
\end{equation}
where $\phi$ is the volume filling factor of the dust aggregate and $\phi_\mathrm{cp} = 0.74$ is the filling factor of the hexagonal-close packing.
They also studied the rebound behavior of cubic latice structures and agglomerates produced by ballistic agglomeration with one (BAM1) and two (BAM2) migrations \citep{ShenEtal:2008} made out of ice particles. It is difficult to compare the former, highly arranged cubic particle with natural grown agglomerates. The latter agglomerates are of interest. The BAM1 and BAM2 agglomerates have coordination numbers of 4 and 6 and volume filling factors of $\sim 0.23$ and $\sim 0.34$, respectively.
The simulations for agglomerates, based on hexagonal packing, by \citet{WadaEtal:2011} have shown that highly porous dust aggregates (low coordination number) stick to each other even if their collision velocity is 22 \metersecond\ (mind that the discrepancy between this high value and the fragmentation threshold of $\sim$1 \metersecond\ is mostly due to the factor of 15 in monomer diameter and the usage of two different materials, i.e. $\mathrm{SiO_2}$ and $\mathrm{H_2O}$ ice). Only if the coordination number reaches values larger than $\mathrm{n_c = 6}$, the dust aggregates (based on closest packing) in the simulations by \citet{WadaEtal:2011} bounce off at intermediate (2.2 \metersecond) and high velocities (22 \metersecond). With Eq. \ref{Eq.nc}, this coordination number corresponds to a volume filling factor of approximately $\phi = 0.37$. In the studies of the two BAM agglomerates, bouncing was also found for BAM2 agglomerates with $n_c=6$, only. For these aggregates no dependency on the velocity was found. \citet{WadaEtal:2011} also found that this critical value for the coordination number is independent of the agglomerate material, because above these values the particles are mechanically locked. This prohibits the dissipation of the collision energy by restructuring of the agglomerate and leads to bouncing.
\citet{WadaEtal:2011} argue that the bouncing observed in experiments might be caused by a preparation-induced compacted surface layer or the overall filling factor of the agglomerates.
While \citet{WadaEtal:2011} can measure the volume filling factor and the coordination number of their simulated aggregates, we can only measure the filling factor and, therefore, depend on a relation between these two values. The relation for the coordination number given by \citet{WadaEtal:2011} (Eq. \ref{Eq.nc}) yields higher coordination numbers than relations for different structures referred to in other publications. An overview about some other relations is given in Fig. \ref{fig.coordination_number}. \citet{Lagemaat:2001} calculated the dependence of the coordination number on the volume filling factor for a randomly packed $\mathrm{TiO_2}$ nano-particle film and found a coordination number of $n_\mathrm{c} = 6$ for a volume filling factor of $\phi = 0.57$. Furthermore, \citet{Antwerpen:2010} gave a summary of different simulations and experiments and according to this paper, the highest coordination numbers (in the range relevant to this work) are achieved by the equation derived by \citet{Rumpf:1958} and the lowest by \citet{Yangetal:2000}. Following this, the critical volume filling factors for bouncing (i.e., corresponding to $n_\mathrm{c} = 6$) are $\phi = 0.48$ and $\phi = 0.65$, respectively. Recent numerical studies by Schmidt and Blum (pers. comm.), who simulated the growth of agglomerates by random ballistic deposition (RBD), where the particles stick at the first contact, and random gravitational deposition (RGD), where particles were allowed to roll and bounce, have also yielded comparable results (triangles in Fig. \ref{fig.coordination_number}).
\citet{SeizingerKley:2013} simulated collisions between microscopic dust aggregates. To produce agglomerates which are as realistic as possible they also used a ballistic aggregation with migration method. An agglomerate is produced by monomers impacting from random directions on the target. After the first contact with the agglomerate the monomer can either stick or migrate to a position where it is in contact with two or three monomers. The selection of the final position has been chosen by three different ways: By selecting the closest position to the impact point (BAM shortest), by selecting a random position (BAM random), or  by selecting the position which is closest to the center of the agglomerate (BAM center). As shown in Fig. \ref{fig.coordination_number}, the three methods lead to different coordination numbers for different volume filling factors. All of these methods lead to coordination numbers, comparable to the other models, but lower than the BAM agglomerates used by \citet{WadaEtal:2011}. This is caused by different calculation methods of the filling factor. \citet{SeizingerKley:2013} used BAM agglomerates which were "sliced out" of a larger agglomerate. Therefore, irregularities on the agglomerate's surface do not decrease the filling factor. Also their agglomerates are considerable bigger than the ones from \citet{ShenEtal:2008}, which also reduces surface effects. This makes the agglomerates more comparable to the particles used, for example, in \citet{WeidlingEtal:2012} or this paper.
In this paper, we will show that bouncing in aggregate-aggregate collisions occurs for lower coordination numbers than those predicted by \citet{WadaEtal:2011}.

In Section \ref{subsection_MPE} we will present our experimental setup, which allows us to study low-velocity collisions between dust agglomerates and their growth under micro-gravity conditions(also see Paper I). We will discuss the properties of the experimental dust agglomerates in Section \ref{subsection_Particles}. Subsequently, in Section \ref{subsection_structure} we will describe the internal structure of our dust aggregates with respect to the discussion above. The results of our collision experiments will be described in Section \ref{section_Results} where we will focus on the transition between sticking and bouncing and the formation of larger dust clusters. Section \ref{section_Discussion} deals with the modeling of the data, and a conclusion of the paper is given in Section \ref{section_conclusion}.

\section{Experimental Setup}\label{section_Setup}
In this Section, we will briefly describe the experimental setup and the dust agglomerates used in our study. A more detailed description of the experimental method can be found in Paper I.

\subsection{Multiple Particle Experiment}\label{subsection_MPE}
\begin{figure*}[ptb]
    \center
    \includegraphics[width=\textwidth]{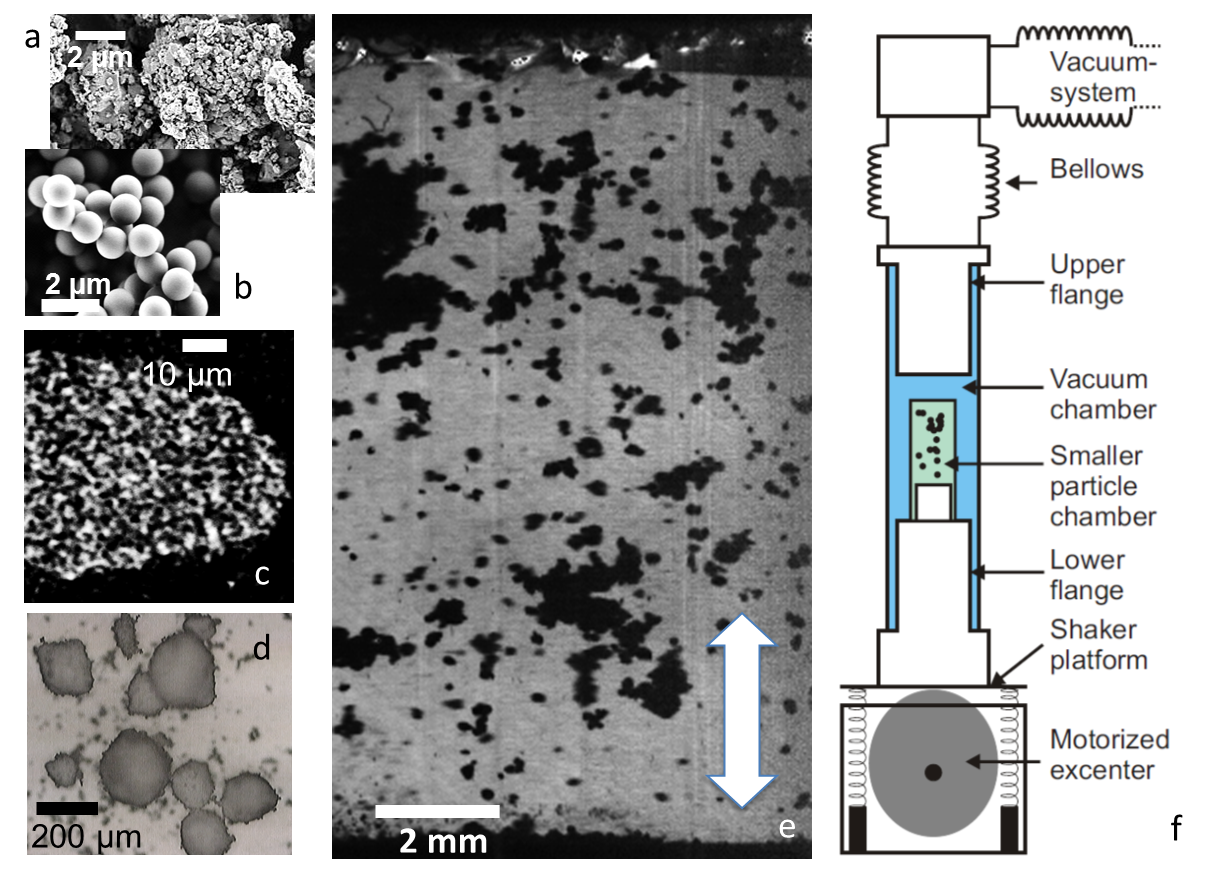}
    \caption{\textbf{a.} SEM image of polydisperse, irregular dust grains with a size distribution of 0.5 - 10 $\mu$m. \textbf{b.} SEM image of monodisperse, spherical dust particles with a diameter of 1.5 $µ$m. \textbf{c.} Computer tomography image of a dust agglomerate consisting of monodisperse $\mathrm{SiO_2}$ particles prior to the experiment. \textbf{d.} Optical microscopy image of sieved dust agglomerates. \textbf{e.} Snapshot from one of the experiments. \textbf{f.} Experimental setup (adapted from Paper I).}
    \label{Fig_DustParticle_Setup}
\end{figure*}

Studies of collisions between dust aggregates at very low impact velocities are difficult to achieve in the laboratory due to the presence of gravity. Therefore, we used the Bremen drop tower to generate a large number of collisions between dust agglomerates under micro-gravity and vacuum conditions. The setup of this experiment is based on the shaking device originally introduced in Paper I (see Fig. \ref{Fig_DustParticle_Setup}f). It consists of a particle container, a shaking mechanism to agitate it and thus generate the collisions, an LED array with diffusor screens for back-light illumination, and a high-speed camera to record the dust-aggregate motion and collisions during the experimental runs. In addition to the setup described in Paper I, the much smaller particles are contained in a small cylindrical glass housing with 1 cm diameter and a height of 1.6 cm. The glass cylinder itself is mounted inside a larger glass vacuum chamber, which is connected to and thus evacuated by the drop tower . To reduce the pressure even more we additionally used a turbomolecular pump. A fine-meshed sieve was attached to the top of the inset container to act as a wall for the particles but enable us to evacuate the smaller volume. As described in Paper I, the vacuum chamber is attached to a platform which can be shaken by an eccentric wheel driven by a motor. With a fixed amplitude of the wheel, the velocity with which the particles are agitated inside the vacuum chamber is changed with the rotation frequency of the motor. Once the experiment is under micro-gravity conditions, the shaker distributes the particles through the container (see Fig. \ref{Fig_DustParticle_Setup}e). The dust aggregates are illuminated by a back-light LED array so that the dust aggregates appear dark in front of a brighter background. To record the experiment, we used a high-speed camera with a resolution of $512 \times 512$ pixels and a repetition rate of 500 frames per second. The camera observes the experiment through a prism, which generates two images separated by a viewing angle of 30 degrees.
\par
In this paper, we analyzed four experimental runs of the experiment (see Table \ref{tab.exp_overview}). We used two dust-analog materials described in the next section in two experiments each. The total duration of each of the four experimental runs was $\sim$9 seconds. For experiments 1 and 2, the shaking frequency was reduced to 50\% of the initial value after 5 seconds. In experiments 3 and 4, the frequency was reduced to 50\% of the initial value after 2 seconds, to 35\% after 5 seconds, and the motor was completely shut off after 8 seconds.

\begin{table*}[tb!]
    \caption{\label{tab.exp_overview}Overview of the performed experiments. The colors in the time table indicate the performance of the shaking mechanism from the start of the experiment (SoE) to the end of the experiment (EoE). \colorbox{gray100}{\color{white}{\textbf{100\%}}} , \colorbox{gray50}{\color{white}{\textbf{50\%}}} , \colorbox{gray35}{35\%} , and \colorbox{gray0}{0\%} .}
    \begin{center}
\begin{tabular}{|c|c|c|c|c|c|c|c|c|c|c|c|c|c|c|}
  \hline
  Experiment & Analog   & Initial    & Initial         & No. of   &  \multicolumn{10}{|c|}{shaker profile}               \\
  number     & material & aver. size & number density  & observed & 0s  & 1s & 2s & 3s & 4s & 5s & 6s & 7s & 8s &9s \\
             &          & [$\mu$m ]  & $[10^9 \rm m^{-3}]$ & collisions &  SoE   &    &    &    &    &    &    &    &    &  EoE \\
  \hline
   1 & polydisperse & 150 & $1.59 $ & 13 & \colorbox{gray100}{\color{white}{}}    &  \colorbox{gray100}{\color{white}{}}  &  \colorbox{gray100}{\color{white}{}}  &  \colorbox{gray100}{\color{white}{}}  &  \colorbox{gray100}{\color{white}{}}  &  \colorbox{gray50}{}  &  \colorbox{gray50}{}  &  \colorbox{gray50}{}  &  \colorbox{gray50}{}  & \colorbox{gray50}{}   \\
   2 & polydisperse & 150 & $1.59$ & 11  & \colorbox{gray100}{\color{white}{}}    &  \colorbox{gray100}{\color{white}{}}  &  \colorbox{gray100}{\color{white}{}}  &  \colorbox{gray100}{\color{white}{}}  &  \colorbox{gray100}{\color{white}{}}  &  \colorbox{gray50}{}  &  \colorbox{gray50}{}  &  \colorbox{gray50}{}  &  \colorbox{gray50}{}  & \colorbox{gray50}{}   \\
   3 & monodisperse & 150 & $1.50$ & 3   &  \colorbox{gray100}{\color{white}{}}   &  \colorbox{gray100}{\color{white}{}}  & \colorbox{gray50}{}   &  \colorbox{gray50}{}  &  \colorbox{gray50}{}  & \colorbox{gray35}{}   &  \colorbox{gray35}{}  &  \colorbox{gray35}{}  &  \colorbox{gray0}{}  &  \colorbox{gray0}{}  \\
   4 & monodisperse & 150 & $1.50$ & 15   &  \colorbox{gray100}{\color{white}{}}   &  \colorbox{gray100}{\color{white}{}}  & \colorbox{gray50}{}   &  \colorbox{gray50}{}  &  \colorbox{gray50}{}  & \colorbox{gray35}{}   &  \colorbox{gray35}{}  &  \colorbox{gray35}{}  &  \colorbox{gray0}{}  &  \colorbox{gray0}{}  \\
  \hline
\end{tabular}
    \end{center}
\end{table*}

\subsection{Characterization of the Dust Aggregates}\label{subsection_Particles}

\begin{figure}[tbp]
    \includegraphics[width=\columnwidth]{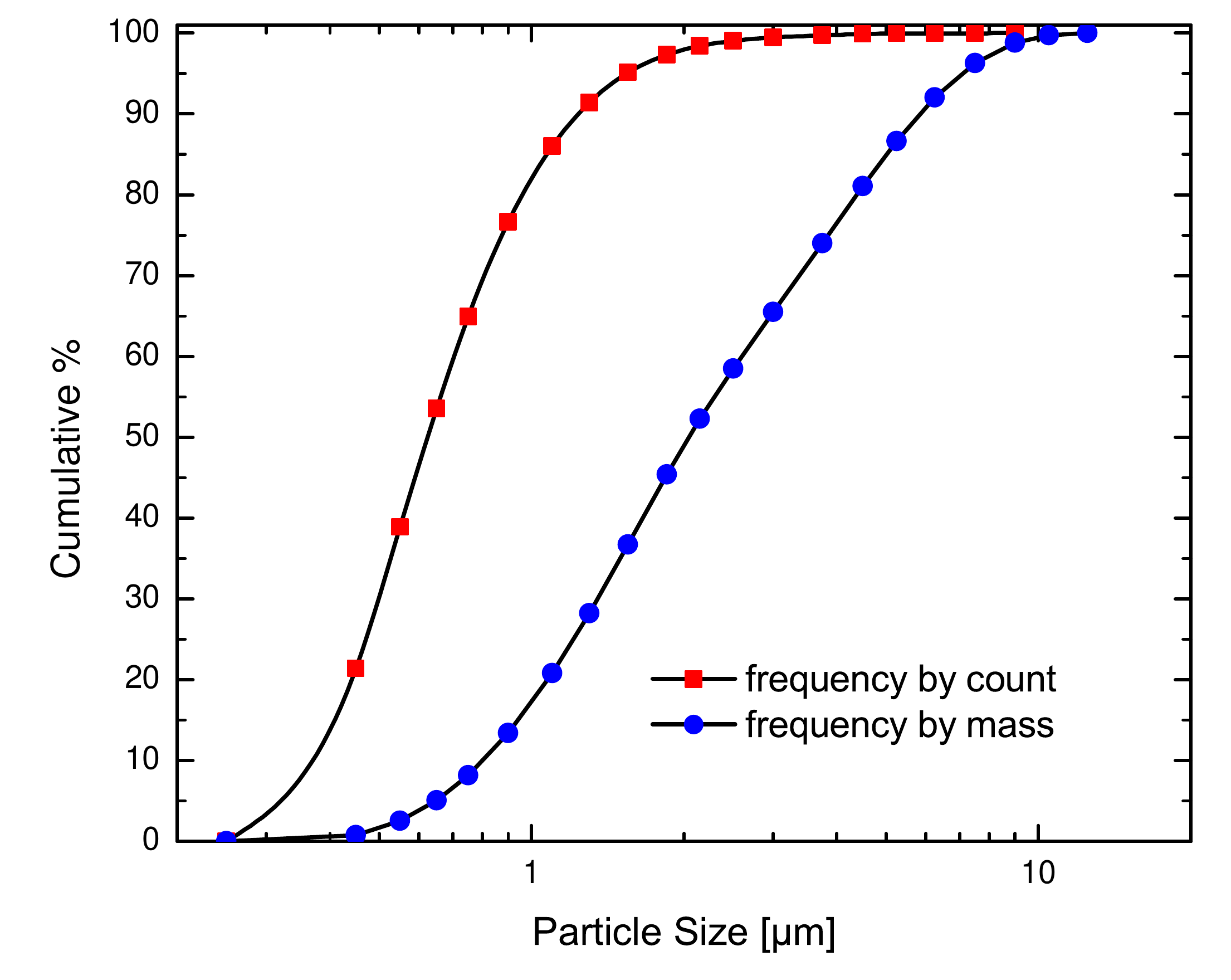}
    \caption{Cumulative size distribution of the irregularly shaped $\mathrm{SiO_2}$ particles. Red squares: cumulative frequency by count of the particles. Blue spheres: cumulative frequency by mass of the particles.}
    \label{Fig_DustSizeDistribution}
\end{figure}

The experiments were conducted with two different dust analog materials as shown in Fig. \ref{Fig_DustParticle_Setup}a and \ref{Fig_DustParticle_Setup}b. The first analog material is the same as used in Paper I\footnote[1]{Manufacturer: Sigma-Aldrich}. It consists of $\mathrm{SiO_2}$ particles with an irregular shape and a wide size distribution, shown in Fig. \ref{Fig_DustSizeDistribution}. The average particle size by number is 0.63 $\mu$m and by mass 2.05 $\mu$m. Approximately 80 \% of the particles are smaller than 1 $\mu$m, while more than 80 \% of the mass is in particles in excess of 1 $\mu$m diameter. The mass density of the monomer grains is 2.6 $\mathrm{g\, cm^{-3}}$. The second material consists of $\mathrm{SiO_2}$ spheres with a monodisperse size distribution and a diameter of 1.5 $\mathrm{\mu m}$\footnote[2]{Manufacturer: micromod Partikeltechnologie GmbH}. The mass density of this material is 2.0 $\mathrm{g\,cm^{-1}}$. Both materials have been used in many earlier experiments and are listed as $\alpha_3$ and $\alpha_4$ in \citet{BlumWurm:2008}.
\par
The dust agglomerates used in our experiments were formed in the dust-storage containers. To prepare the dust agglomerates, we sifted the dust through two sieves with a mesh size of 250 $\mathrm{\mu}$m (upper sieve) and 100 $\mathrm{\mu}$m (lower sieve). The selected dust agglomerates have an average size of $150 \pm 60$ \micrometer  (standard deviation) and have a roughly spherical or ellipsoidal shape (see Fig. \ref{Fig_DustParticle_Setup}d).
\par
The volume filling factor of the sieved dust aggregates was derived in Paper I for several specimen by measurements of the dust-aggregate masses with a high precision balance and their respective volumes with optical methods. In Paper I, an average volume filling factor of $ \phi = 0.35 \pm 0.05$ was determined for mm-sized dust aggregates with masses between 0.1 and 30 mg. The dust aggregates studied in this paper are considerably smaller so that the determination of the mass and volume of individual dust aggregates was not feasible with this method.\\
However, to get a more accurate measurement of the volume filling factor and also to study the inner structure of our dust aggregates (for those with monodisperse monomers), we used nano-CT images obtained with a SkyScan 2011 scanner with a voxel size of 0.3 $\mathrm{(\mu m)^3}$. We analyzed one dust aggregate with a triaxial ellipsoidal shape and semi-major axes of approx. 150 \micrometer, 100 \micrometer, and 40\micrometer, prior to the experiments. We calculated the volume filling factor by measuring the average gray value in the images and normalized this to the maximum value in the sample, which represents the density of solid $\mathrm{SiO_2}$. The difficulty of this method is to distinguish between the gray-value of the bulk material and artifacts caused by the reconstruction of the tomography images. To calculate the gray value for $\mathrm{SiO_2}$, we measured the maximum gray values inside of 50 monomer grains. At the given resolution, each particle is supposed to have a diameter of 5 voxels. Due to reconstruction and scattering effects, the particles seem to be a little larger, with decreasing brightness towards the rim. However we can assume that the innermost voxel represents the gray value of the bulk material. This method leads to a volume filling factor of $\phi = 0.37^{+0.06}_{-0.05}$, in agreement with the method used in Paper I.
While the volume filling factor is directly measurable, the coordination number was identified by \citet{WadaEtal:2011} to be the critical value for bouncing.
According to Eq. \ref{Eq.nc} our agglomerates correspond to a coordination number of $n_\mathrm{c} = 5.9$, which is close to the critical value of $n_\mathrm{c} = 6$. However, using the relations between coordination number and volume filling factor by \citet{Lagemaat:2001}, \citet{Rumpf:1958} and \citet{Yangetal:2000}, the measured volume filling factor rather corresponds to coordination numbers of $n_\mathrm{c} = 3.76$, $n_\mathrm{c} = 4.92$, and $n_\mathrm{c} = 3.56$, respectively. These values are significantly lower than the critical value derived by \citet{WadaEtal:2011}.

\subsection{Inner Structure of the Dust Aggregates}\label{subsection_structure}
\begin{figure}[ptb]
    \includegraphics[width=\columnwidth]{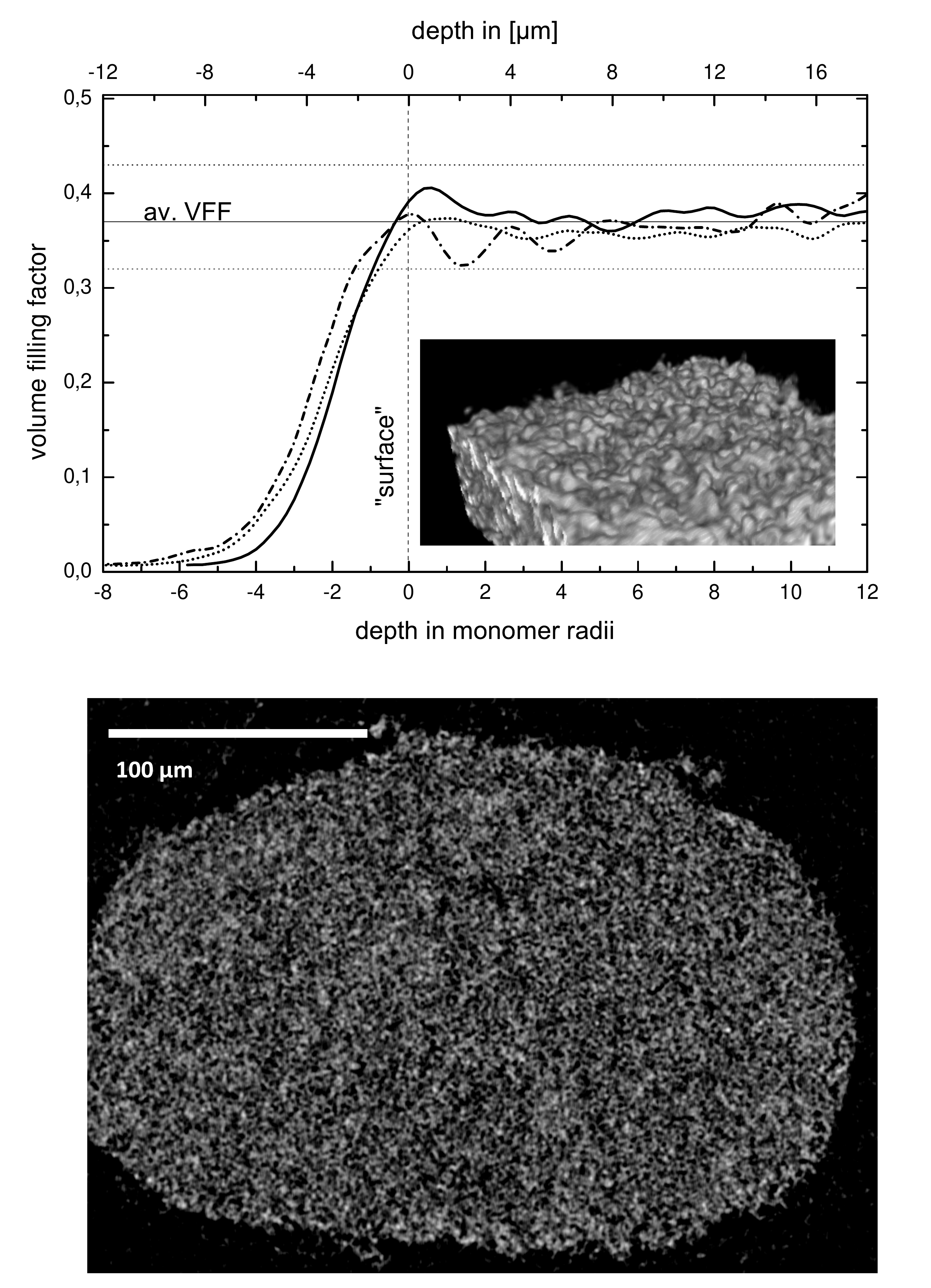}
    \caption{\textit{Top:} Measurement of the volume filling factor close to and below the surface of the analyzed dust agglomerate. The three curves represent the measurements for three different positions on the surface (\emph{solid curve:} 245 px $\times$ 265 px, \emph{dotted curve:} 400 px $\times$ 257 px, \emph{dash-dotted curve:} 103 px $\times$ 142 px). The inset shows a 3D reconstruction of one of the analyzed regions. \textit{Bottom:} NanoCT image of a cross-section through the dust agglomerate with semi-major axes of 150 $\mu$m, 100 $\mu$m, and 40 $\mu$m.}
    \label{fig.Rim_thickness_MEDEA}
\end{figure}

In Section \ref{section_Introduction}, we discussed that the bouncing of dust agglomerates is currently debated between theorists and experimentalists. The question has risen whether bouncing, which has been observed in experiments but not in numerical collisions, is an artifact due to a compacted shell of the dust agglomerates. This compact mantle could be caused by the handling process or by the production method of the dust aggregates as argured by \citet{WadaEtal:2011}. We therefore analyzed the rim structure of the nano-CT scanned dust aggregate at three different positions. Figure \ref{fig.Rim_thickness_MEDEA} shows the density curve close to the surface and down to a depth of 18 $\rm \mu m$, which corresponds to 12 monomer diameters. The density at the dust agglomerate's surface is close to the average value measured in the center of the dust agglomerate. Our measurements show a slight oscillation in the density profile. The variation in the volume filling factor is on the order of $\Delta \phi = \pm 0.05$. With Eq. \ref{Eq.nc} this corresponds to a deviation from the mean coordination number of $\Delta n_c = \pm 0.4$. We assume that the measured oscillation indicates a spatially regular structure close to the surface.

\subsection{Determination of the Collision Properties}\label{subsection_VeloDeterm}

The main scope of our experiments was to study the outcome of collisions between sub-mm-sized dust agglomerates. We used a semi-automatic tracking program for individual dust agglomerates, which enables us to measure the collision velocities and dust-aggregate masses for identified collisions. However, only a small fraction of the actual number of collisions could be observed, due to optical-depth and identification problems. We are not aware of any preferred selection of a certain collision type so that the observed collisions should be treated as a random representative (but incomplete) sample.
\par
After a collision was identified, the trajectories of the colliding dust aggregates were tracked before the collision for as long as possible. From these trajectories, we calculated the collision velocities. It has to be taken into account that the dust agglomerates were only followed in two dimensions, due to the large number of individual dust agglomerates in both projected fields of view and the consequential impossibility to unambiguously identify a given dust aggregate on the other projection. However, \citet{WeidlingEtal:2012} showed that this restriction leads to a statistical error of only 13\% in the collision velocity. We also neglected the rotation of the dust aggregates, which was clearly present but not measurable. We also checked the particle trajectories for any sign of acceleration or deceleration, due to gas drag or electric charges. All measured trajectories were linear in space and time within the accuracy of our measurements.
\par
The mass of the dust agglomerates was estimated from the average projected area when the particle was tracked. Assuming a spherical shape of the dust aggregates, this leads to a dust-aggregate mass of
\begin{equation}
\label{Eq.Mass}
    M = \frac{4}{3 \sqrt{\pi}} \phi \varrho_\mathrm{SiO_2} \hat{A}^{3/2}
\end{equation}
where $\phi=0.37$, $\varrho_\mathrm{SiO_2}$, and $\hat{A}$ are the volume filling factor of the dust aggregate, the mass density of the monomer grains, and the average projected area of the dust aggregate  while the particle is tracked, respectively.

\section{Results}\label{section_Results}
\begin{figure*}[!p]
    \center
    \includegraphics[width=\textwidth]{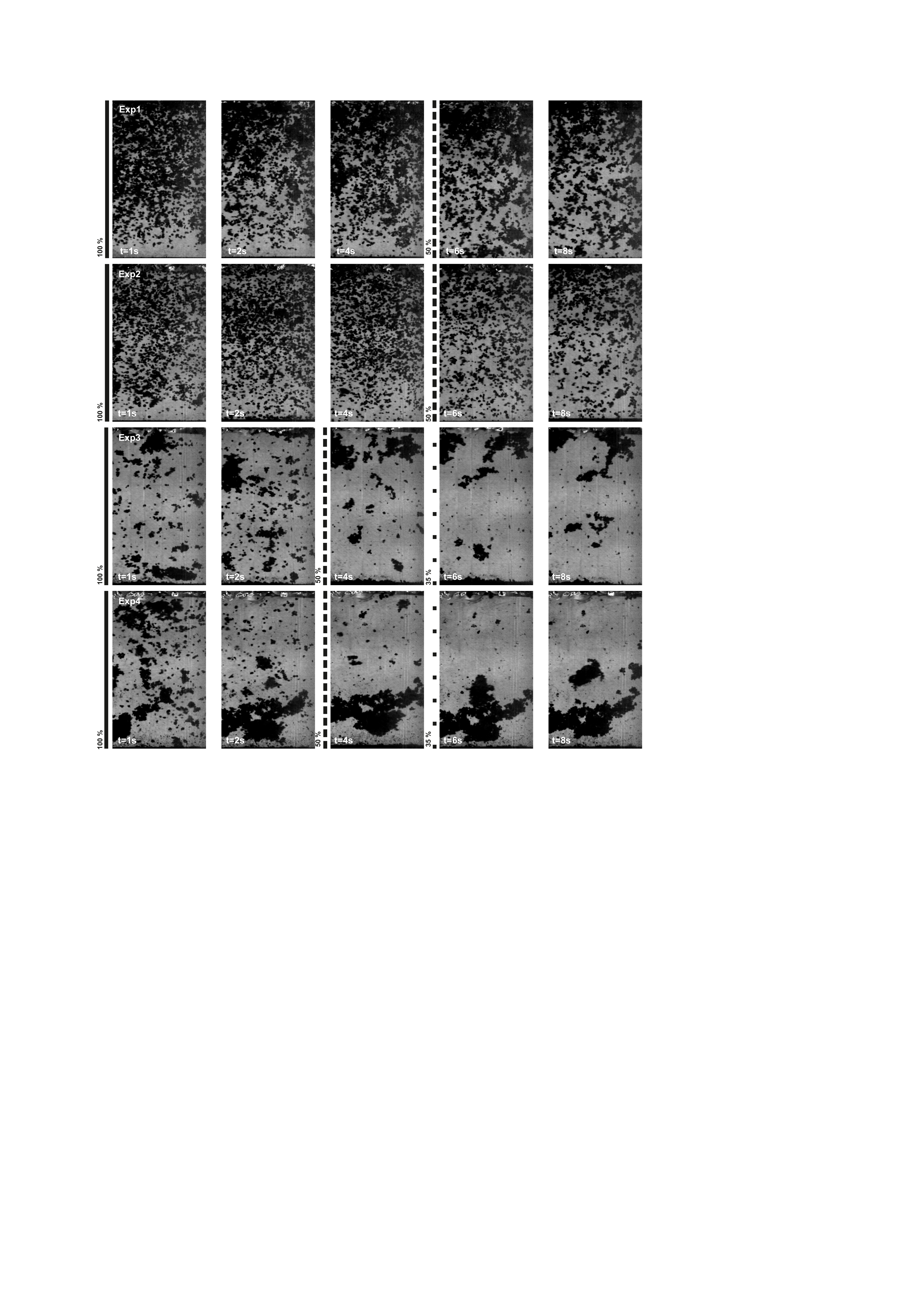}
    \caption{Snapshots of the four experiments at different times. The first two columns show the particles during the deagglomeration phase (100\% shaker speed). The speed of the shaker is indicated by bars next to the images. For Exp1 and Exp2, the shaking frequency was reduced to 50\% of the initial value after 5 seconds. In Exp3 and Exp4, the frequency was reduced to 50\% of the initial value after 2 seconds, to 35\% after 5 seconds, and set to 0 after 8 seconds. The right column shows the experiments at their final states.}
    \label{Fig_Snapshots}
\end{figure*}

\begin{figure*}[ptb]
    \center
    \includegraphics[width= \textwidth]{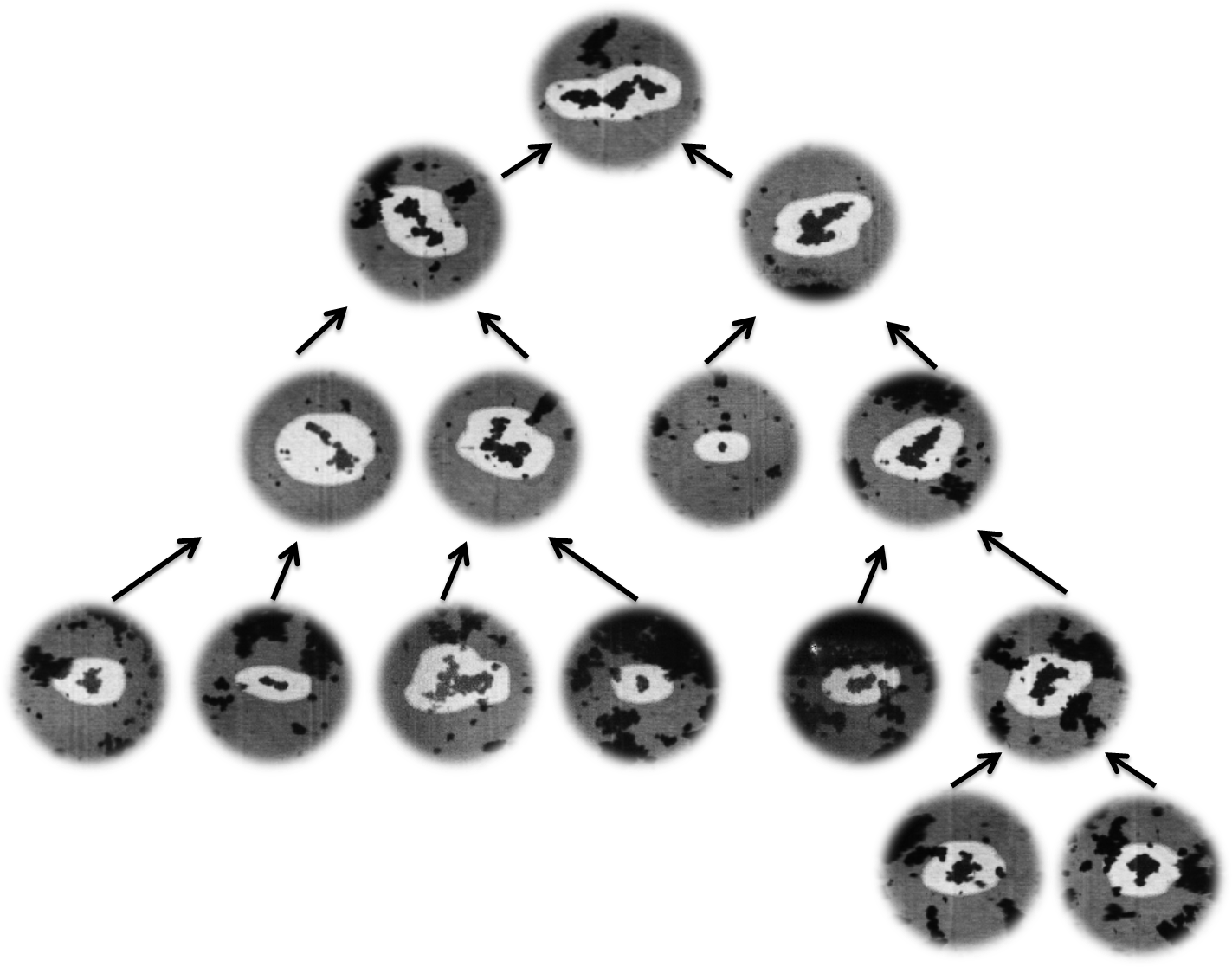}
    \caption{Illustrations of the sequence of hit-and-stick collisions between clusters of dust agglomerates in Exp3 leading to a fractal cluster. The collision velocities range between 1.6 $\rm cm~s^{-1}$ and 5.9 $\rm cm~s^{-1}$. (A movie of Exp3 is available in the online journal).}
    \label{Fig_Particle_Tree}
\end{figure*}

Figure \ref{Fig_Snapshots} shows the temporal evolution of the four dust-aggregate dispersions during their respective experiments. The first column shows a snapshot after 1 second. The agglomerates made of polydisperse dust (Exp1 and Exp2) are well distributed throughout the experiment volume. In contrast, the agglomerates made of monodisperse dust (Exp3 and Exp4) are only partially deagglomerated. A large fraction of the mass in Exp3 and Exp4 is still bound within a few large clusters, which we were not able to deagglomerate with our setup and the given experiment time. We call an agglomerate a cluster if it consists out of several agglomerates (like the large agglomerates in Fig. \ref{Fig_Particle_Tree}) from the initially sieved particles (see Fig \ref{Fig_DustParticle_Setup} d).

As we will show in Section \ref{subsection_sticking_of_fracatls}, individual particles are more likely to stick to clusters than to one another. Therefore, the single particles in Exp3 and Exp4 get depleted within a short time after the shaking frequency and, thus, also the average velocity has been reduced. In Exp3 we were able to observe the formation of a large and elongated cluster out of smaller dust agglomerates and clusters. This object has fractal-like characteristics and was formed in a succession of collisions as shown in Fig. \ref{Fig_Particle_Tree}. On the other hand, the polydisperse dust particles did not form very large clusters within the experiment duration. A reason for this might be that the shaking frequency was not reduced below 50\% of the initial value. This caused the optical depth in these experiments to be higher than in Exp3 and Exp4 and made it difficult to identify individual collisions between dust agglomerates consisting of irregular $\rm SiO_2$, albeit there should be many. For the experiments with monodisperse dust, the number of detectable collisions between individual sub-mm-sized dust aggregates is also limited, due to their low number density and due to the presence of the large clusters, which collide with the remaining particles more frequently due to their larger cross section.

In Section \ref{subsection_Sticking_monomers}, we will describe the outcome of the observed collisions between pairs of individual dust agglomerates. The properties of the cluster structure from a sequence of sticking collisions will be studied in Section \ref{subsection_formation_of_fractals}. A comparison between collisions of individual dust aggregates and those of clusters of aggregates will be performed in Section \ref{subsection_sticking_of_fracatls}.

\subsection{Sticking and Bouncing of Single Sub-mm-sized Dust Aggregates}\label{subsection_Sticking_monomers}
\begin{figure}[ptb]
    \includegraphics[width=\columnwidth]{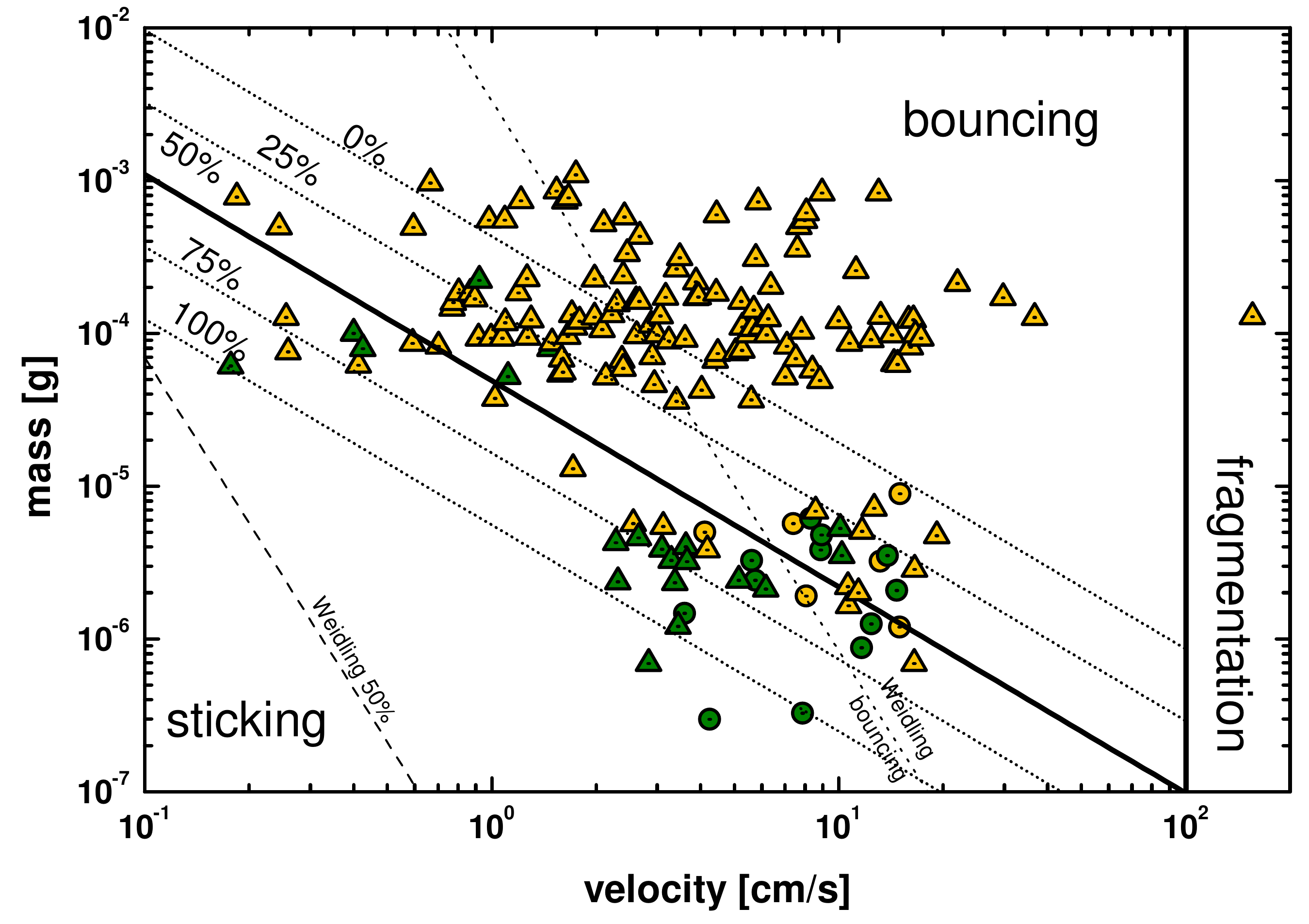}
    \caption{Outcomes of the collisions between individual sub-mm-sized dust aggregates. The axes denote to the collision velocity and the mass of the smaller collision partner. Triangular and circular symbols represent collisions between dust aggregates consisting of polydisperse and monodisperse $\rm SiO_2$ grains, respectively. The upper cloud of data points are the collisions studied in Paper I. The long-dashed and short-dashed lines represent the 50\%/50\% and 0\%/100\% sticking/bouncing probabilities from Paper I (see Table \ref{tab.CVaR}). A new power law at 50\%/50\% sticking/bouncing probability was calculated to divide collisions which preferably lead to sticking for lower and bouncing for higher velocities. The fit is based on the results from this work (low masses) and Paper I (high masses). The dotted and solid lines indicate the continuous transition between perfect sticking (100\%) and perfect bouncing (0\%).}
    \label{Fig.newmodel}
\end{figure}

We analyzed a total number of 42 collisions between individual sub-mm-sized dust aggregates. The collision velocities were between 2.3 $ \rm cm~s^{-1}$ and 16.6 $ \rm cm~s^{-1}$ and the dust-aggregate masses were between $3.0 \cdot 10^{-7} $g and $9.3 \cdot 10^{-6}$g. The collision partner had approximately the same size, with more than 85\% of the particles having a mass ratio smaller than 5. The outcome of these collisions is presented in Fig. \ref{Fig.newmodel}, in which the green and yellow data points denote sticking and bouncing collisions, respectively, and the long-dashed and short-dashed lines represent the 50\%/50\% and 0\%/100\% sticking/bouncing probabilities from Paper I, of which the parameters are given in Table \ref{tab.CVaR}. Triangular and circular symbols represent collisions among dust aggregates consisting of polydisperse and monodisperse $\rm SiO_2$, respectively. A collision was identified as sticking when the particles sustained their contact for several frames. In particular, individual particle rotation after the collision illustrates that the sticking was impermanent, whereas a rotation of both colliding aggregates about a joint axis for an extended period of time indicates sticking. The observation time after a collision was limited by either a collision with a wall, a cluster covering the sticking agglomerates or the end of the experiment. The upper cloud of data points in the mass range around $10^{-4}$ g was taken from Paper I. As described earlier, our experiments were performed with two different dust samples. However, we found no strong evidence that the transition between sticking and bouncing depends on the monomer size distribution and morphology which is contrary to our observation that larger agglomerates only form in experiments conducted with the monodisperse dust partricles. This will be diskussed in Section \ref{section_Discussion}.
It is obvious from Fig. \ref{Fig.newmodel} that collisions among sub-mm-sized dust aggregates lead to sticking at velocities higher than predicted by Eq. 8 in Paper I. Sticking also occurred at velocities for which pure bouncing was expected.

\subsection{Structures of Large Dust-Agglomerate Clusters}\label{subsection_formation_of_fractals}

In Section \ref{subsection_Sticking_monomers}, we showed that many of the observed collisions lead to sticking in the velocity range investigated. However, the final result of the growth process at the end of the experiment can be different. In Exp3 and Exp4, we found that at the end of the experiment almost all initial dust aggregates were bound in large clusters (some of these clusters had survived the deagglomeration phase at the beginning of the experiments). We also observed the formation of elongated and highly porous structures (see Fig. \ref{Fig_Particle_Tree}). The formation of these fractal-like dust agglomerates has also been observed in later drop-tower experiments (Brisset et al., in prep). Measuring the fractal dimension of these large clusters is difficult, due to the finite initial dust-aggregate size and the limited resolution of the images. A rough estimation of the value of the fractal dimension can be derived by the box-counting method \citep{Falconer:1990}. For this method, the projected image of the particle is placed inside a two-dimensional box of the length l. This box is divided in $\epsilon$ grid cells with length $\mathrm{l}/\epsilon$. The number of grid cells is increased step by step. In each calculation step, the number N of boxes which cover a part of the agglomerate is counted. The fractal dimension is calculated from the relation $\mathrm{N}(\epsilon)=\epsilon^{-D_f}$ or taken from the slope of the linear function in a log (N) / log (1/$\epsilon$) plot. However, this method should only be treated as a rough estimation. The finite size of the $\sim$ 150 $\mu$m agglomerates limits the resolution of the box-counting method. For $\mathrm{D_f} \le 2$ the value obtained from a projection of an agglomerate is comparable (but in some cases smaller) to the real (i.e. three-dimensional) value \citep{Nelsonetal:1990}. In our case the finite size of the initial dust aggregates requires an additional estimation of the error of our method. We applied the box-counting-method to every image of the agglomerate, which corresponds to images from different angles due to the rotation of the aggregates. Therefore, the value of the fractal dimension changes from image to image. These different perspectives allow us an estimation of the error of the method. The average value in the right image is $\mathrm{D_f}=1.58 \pm 0.03$ (more "side views") and $\mathrm{D_f}=1.71 \pm 0.02$  in the left image (more "head-on views"). The average of all images from both frames, considering the different number of available images in both projections, leads to a value $\mathrm{D_f}=1.63 \pm 0.07$.
It must be noticed that there is also a systematic error due to the two-dimensional projection. \citet{Nelsonetal:1990} calculated that the deviation depends on the observation scale (the ratio of the size of the building blocks and the size of the fractal) and the real value of $\mathrm{D_f}$ (Eq.13 and Fig.2 in \citet{Nelsonetal:1990}). The error is negligible for values close to 2 and can be on the order of $ \mathrm{\Delta D_f=}0.1$ for fractal dimensions of $ \mathrm{D_f \approx 0.1}$. If we assume that the size of a typical individual monomer-agglomerate is on the order of 5 px and the analyzed fractal of 100 px, a measured fractal dimension of $\mathrm{D_f = 1.63}$ would correspond to a real value of $\mathrm{D_f = 1.73}$.
For the more compacted clusters, this method loses accuracy. Irregularities with sizes on the order of an individual dust aggregates on the surface of the projected cluster lead to lower values than expected from the images. Average values vary between $\mathrm{D_f = 1.63}$ and $\mathrm{D_f = 1.82}$ with an error that can be on the order of $\mathrm{\Delta D_f = 0.08}$, at which we would expect values closer to 2.\\
The low fractal dimension for the elongated structures is in agreement with hit-and-stick cluster-cluster collisions \citep{BlumEtal:1999,KrauseBlum:2004,PaszunDominik:2006}. This very low value is an indication that the threshold velocity between sticking and bouncing of clusters of dust aggregates is (well) above the experimental velocities.

\subsection{Sticking and Bouncing of Clusters of Dust Aggregates}\label{subsection_sticking_of_fracatls}
In Section \ref{subsection_Sticking_monomers}, we focused on collisions between individual sub-mm-sized dust agglomerates. However, due to the succession of sticking collisions, we expect the dust agglomerates in protoplanetary disks to possess a more complex morphology than spherical. Therefore, collisions between clusters of dust agglomerates, which formed during the runtime of the experiment, are also of interest. We measured the relative velocities for collisions between dust-agglomerate clusters as well as for impacts of single dust aggregates into larger clusters.
\par
In the case of clusters of dust agglomerates the mass determination is difficult. We estimated the mass of the fractal-like objects by summing up the masses of all dust aggregates that were captured during the growth process (see Fig. \ref{Fig_Particle_Tree}). For each of the captured elongated clusters, we modeled the structure from the different images available, by approximating the shape of the cluster with ellipsoids. The size of these "building blocks" was estimated from the available images of the cluster. The mass estimation is the sum of these ellipsoids. In the case of the more compact clusters, we based the mass estimation on the projected area of the clusters and calculated the mass by Eq. \ref{Eq.Mass}. We likely overestimated the masses of these objects, but should still be on the right order of magnitude.
\par
\begin{figure}[ptb]
    \includegraphics[width=\columnwidth]{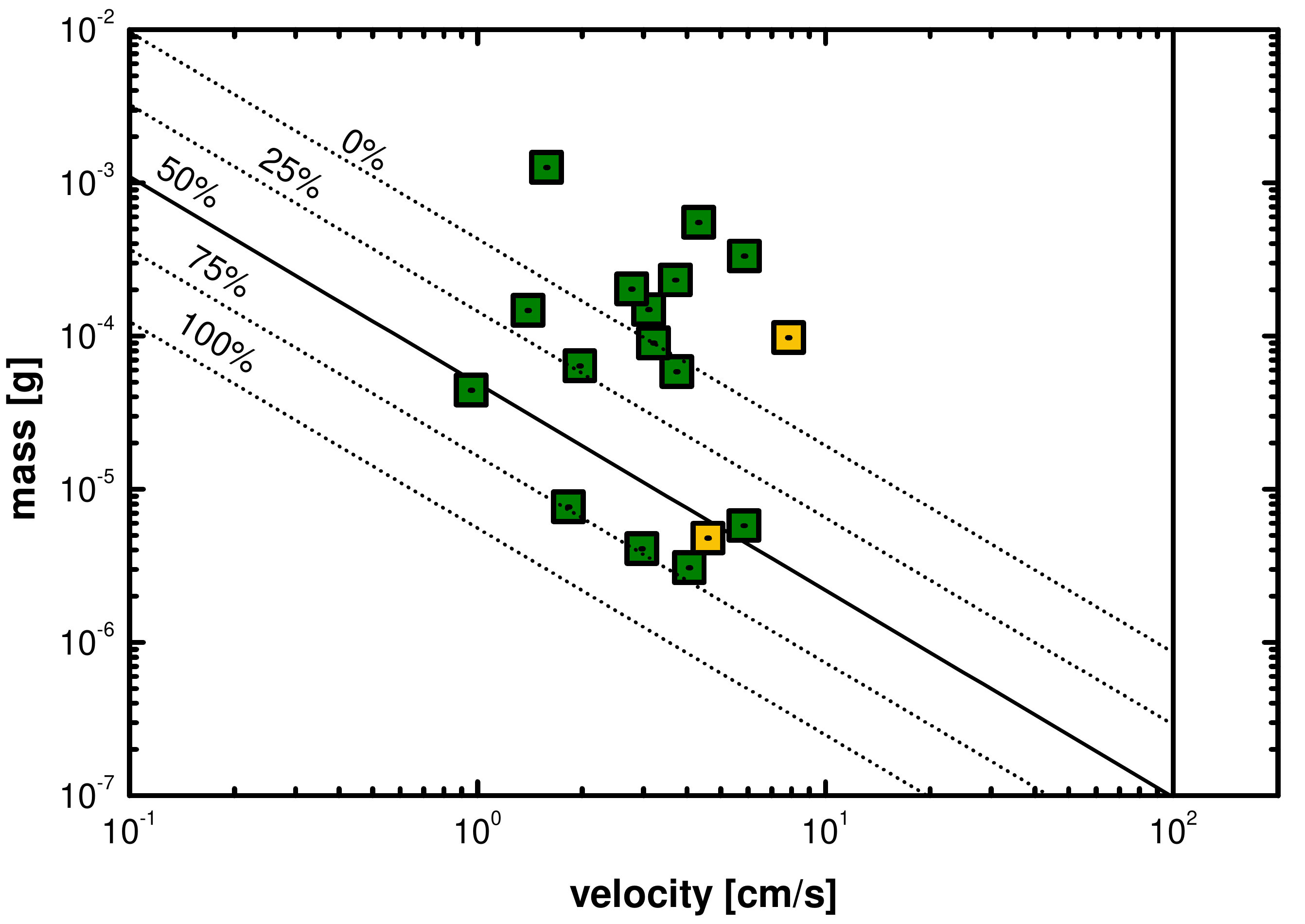}
    \caption{\label{DustClusters}Outcomes in collisions between clusters of sub-mm-sized dust aggregates consisting of monodisperse $\rm SiO_2$ grains. The masses are those of the smaller cluster or, in the case of the five data points below $10^{-5}~\rm g$, individual sub-mm-sized dust aggregates. The dotted and solid lines are the same as in Fig. \ref{Fig.newmodel} and indicate the continuous transition between perfect sticking (100\%) and perfect bouncing (0\%) in collisions between individual sub-mm-sized dust aggregates.}
\end{figure}

The collision outcomes of the cluster impacts of Exp3 and Exp4 are shown in Fig. \ref{DustClusters} in the same way as in Fig. \ref{Fig.newmodel}. The green and yellow data points indicate sticking and bouncing collisions, respectively. The ordinate denotes the mass of the smaller agglomerate or cluster. The mass range of the larger collision partners was between $4.4 \cdot 10^{-5}$g and $1.3 \cdot 10^{-3}$g. The five collisions with dust-agglomerate masses below $10^{-5}~\rm g$ are impacts of individual dust agglomerates into clusters of agglomerates. All other impacts are between clusters of dust agglomerates.
\par
Compared with the data of the collisions among single dust aggregates (see Fig. \ref{Fig.newmodel}), we find many more sticking events than predicted by the 50\%/50\% sticking/bouncing line (solid line in Fig. \ref{DustClusters}). Within the range of our experimental data we were not able to see a clear onset of a sticking / bouncing transition for clusters. It is obvious that the looser structure of the clusters favors sticking, due to a better energy dissipation.

\section{Discussion}\label{section_Discussion}
The experiments in this paper were performed with dust aggregates smaller, but structurally comparable to those used in Paper I. We were able to confirm a transition between sticking and bouncing over a velocity range. However, a comparison with the transition curve and width in Paper I showed that their values (long and short-dashed lines in Fig. \ref{Fig.newmodel}) do not agree with the results of this work. This disagreement is caused by the fact that the dust aggregates in Paper I had a rather narrow size distribution so that a deduction of the mass-velocity relation was not feasible. Thus, Paper I used the contact model by \citet{ThorntonNing:1998} and its slope, which obviously is in contradiction with our new data.
\par
Now, we are in the situation that the combination of the data of Paper I and the data presented here spans more than three orders of magnitude in mass so that we can try and fit a power-law relation between the dust-aggregate mass $m$ and the transition velocity $v$ from sticking to bouncing, i.e.,
\begin{equation}
    \frac{m}{\mathrm{1~g}}  = a \cdot \left (\frac{v}{\mathrm{1~cm/s}} \right )^b ,
    \label{Eq.0.5sticking}
\end{equation}
with $a$ and $b$ being free fit parameters. Due to the qualitatively different evolution of the experiments with different morphology of the individual dust grains (see Fig. \ref{Fig_Snapshots}), we additionally consider a subset of data from experiments conducted with irregular dust only. We performed the fitting procedure with the logarithms of mass and velocity and considered the transition velocity as the 50\%/50\% sticking/bouncing threshold. To calculate the power law, we used six different methods. For all fits, we minimized the quadratic and linear deviation as well as the number of data points on the wrong side of the fit function. For those three approaches, we minimized the error with and without the boundary condition that the number of false data points on both sides of the fit function is equal.
\par
In Table \ref{tab.CVaR} we give an overview of the results of the different methods. Values in parentheses were calculated by using only data from experiments with irregular dust particles. To interpret these results, we used the conditional value at risk (CVaR) or expected shortfall \citep{Hull:2007}, which is often used in finance mathematics as a measure to evaluate the risk of a portfolio. In our case, the CVaR is defined as the mean deviation in velocity of the q\% collisions which deviate the most from the 50\% line. Due to the small number of collisions deviating from our power laws, we used q=20\%. Based on this criterion, the least squares and least linear deviation method and the requirement of a symmetric number of false data points on both sides of the curve as well as the least squares method with asymmetric false data points possess the lowest CVaR.
The least-square and least-linear methods give rather similar exponents ($b \sim -4/3$) and prefactors ($a \sim -4.16...-4.79$) which are, within the accuracy of our measurements and taking the small number of data points into account, in good agreement. Some of the calculation methods lead to several solutions within small derivations in $a$ and $b$. For these cases we give an average value of the parameters and marked them with an asterisk in Table \ref{tab.CVaR}.
\par
We also calculated the threshold power law for the collisions from Paper I and only the results from the experiments with polydisperse dust. This leads to comparable values for the prefactors $a$. However, the power-laws differ slightly from the values obtained with the full data set but do not change the overall picture (see Table \ref{tab.CVaR}).

\begin{table*}[ptb]
    \caption{\label{tab.CVaR} Fit results of the six different methods used to calculate the transition between bouncing and sticking for monodisperse and polydispers dust samples. Values in parentheses are only based on the polydisperse sample from this work and \citet{WeidlingEtal:2012}. Calculations marked with asterisks represent the average of several similar solutions. For details, refer to the text.}
    \begin{center}

\begin{tabular}{|c|c|c|c|c|}
  \hline
  Fit        & Exponent  & Constant    & Tran-  & CVaR \\
  method     &  b        & a           & sition  &[log($\rm cm\,s^{-1}$)] \\
             &           &             & width $\tau $ &  \\
  \hline
  \multicolumn{5}{|c|}{Asymmetric number of false data points}    \\
  \hline
   Least squares         & -1.34 (-1.13)     & $10^{-4.34}$ ($10^{-4.79}$)&  -0.72 (-0.44) & 0.41 (0.86) \\
   Least linear dev.     & -1.36 (-1.13)     & $10^{-4.42}$ ($10^{-4.79}$ ) & -0.78 (-0.45) & 0.46 (0.88)\\
   Least number of       & -0.94 (-0.93)      & $10^{-4.87}$ ($10^{-4.88}$ ) & -0.40 (-0.38)& 1.12 (1.15)\\
   false data points$\ast$    &            &              &        &      \\
 \hline
  \multicolumn{5}{|c|}{Symmetric number of false data points}    \\
  \hline
  Least squares$\ast$                &   -1.35 (-1.48)   &   $10^{-4.31}$ ($10^{-4.33}$)     &    -0.71 (-0.71)    & 0.39 (0.41) \\
  Least linear dev.$\ast$           &  -1.37  (-1.47)   &  $10^{-4.31}$ ($10^{-4.16}$ )      &   -0.73 (-0.69)    & 0.39 (0.45)\\
  least number of           &   -1.30 (0.91)   &  $10^{-4.46}$ ($10^{-4.30}$ )      &   -0.80 (-0.43)    & 0.51 (0.72)\\
  false data points$\ast$     &            &              &        &      \\
  \hline
  \multicolumn{5}{|c|}{Values from Paper I}    \\
  \hline
             &   -3.6   &  $10^{-7.76}$     &        &  \\ 
  \hline
\end{tabular}
    \end{center}
\end{table*}

\begin{figure}[ptb]
    \includegraphics[width=\columnwidth]{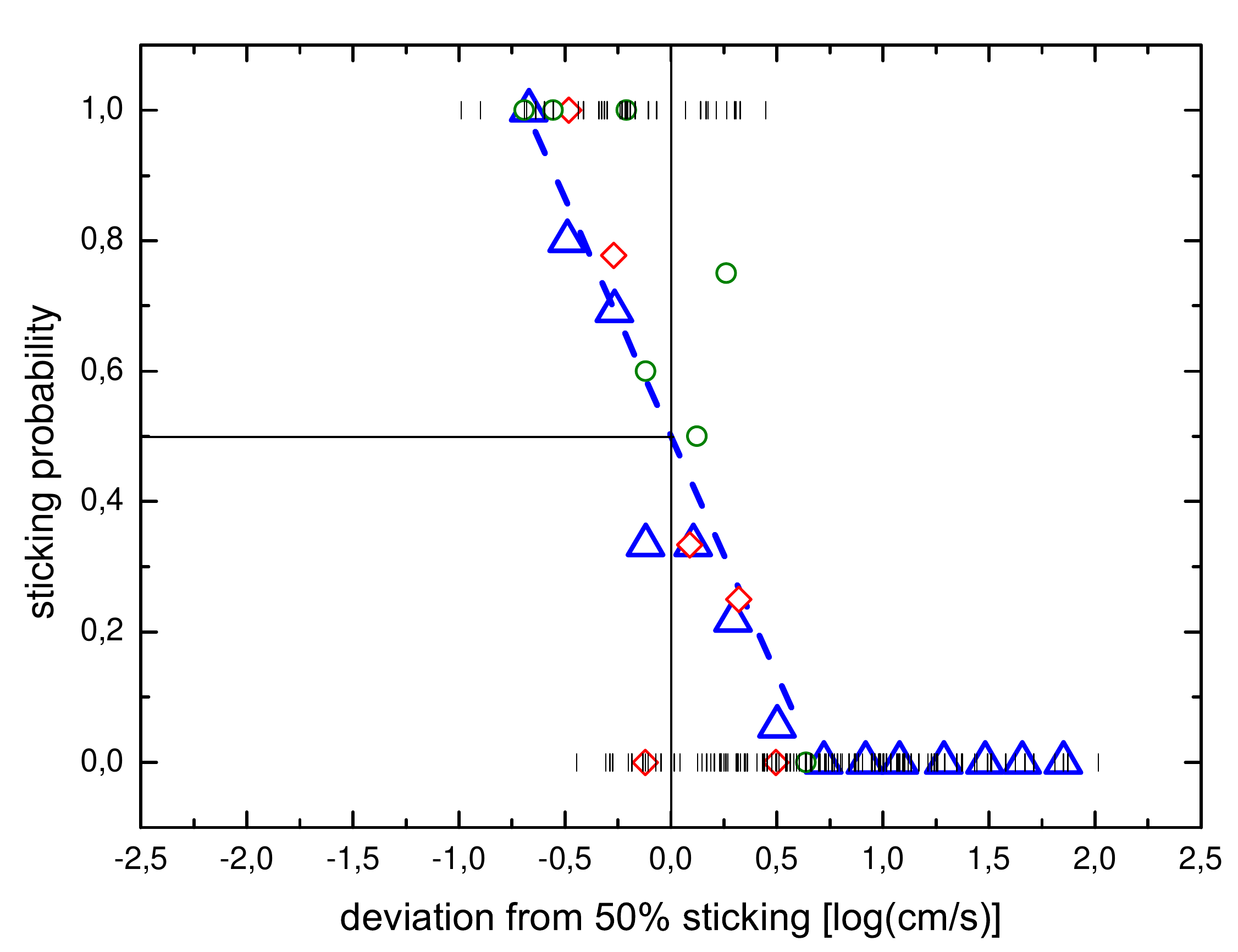}
    \caption{\label{P_Transition} Calculation of the sticking probability based on the data from this work and Paper I. The abscissa shows the logarithm of the deviation from the best fitting power law (least linear dev., symmetric) for the 50\%/50\% sticking/bouncing threshold. Plotted are the results of the individual collisions (black bars) and the sticking probability within equidistant logarithmic velocity bins for both materials (blue) and only the monodisperse (green dots) and polydisperse (red diamonds) from this work. To describe the transition from sticking to bouncing, we applied the linear function shown in Eq. \ref{eq_transition}. The 50\%/50\% sticking/bouncing position was fixed by Eq. \ref{Eq.0.5sticking}.}
\end{figure}

In the same fashion as in Paper I, we assume a logarithmic sticking-probability distribution in velocity so that all threshold lines are parallel. However, we were not able to treat the dust-aggregate masses ranging over more than three orders of magnitude as identical. For each sticking and bouncing collision, we therefore calculated the deviation in velocity according to the power law computed above. As shown in Fig. \ref{P_Transition}, we then used logarithmic equidistant velocity bins to fit a linear function for the probability distribution.
 \begin{equation}
    P = \tau \cdot \Delta + 0.5 ,
    \label{eq_transition}
 \end{equation}
where $P$ is the sticking probability between 0 and 1, $\Delta$ is the logarithmic deviation in velocity from the P=0.5 power-law and $\tau=-0.73$ the slope of the function (least linear deviation and symmetric fit method). The transition function $P(\Delta)$ is plotted in Fig. \ref{Fig.newmodel}. The slight disagreement between the data points and the fit curve is caused by the fixed value for the 50\%/50\% sticking/bouncing threshold. For comparison we also show the sticking probability, within the velocity bins, for our experiments with only monodisperse (green dots) and polydisperse (red diamonds) dust. Considering the smaller number of collisions, these data subsets are still described by the fit derived for all data, even though the agglomerates made out of monodisperse dust stick at slightly higher velocities.

The new power law for the 50\%/50\% sticking/bouncing transition is less steep (slope $\sim -4/3$) than the one given by the model of \citet{ThorntonNing:1998}, with a slope of -18/5, used in Paper I. Small dust aggregates stick at higher velocities than predicted by Paper I. This shows the limits of the previous contact model, which is based on static contacts between solid spheres. Moreover, the transition from perfect sticking to bouncing is much narrower and the reason is the lack of data for sticking collisions in Paper I. It is obvious for the two clouds of data that the transition can be better determined when including the small-aggregate collisions.
\par
The impact of this new transition curve between sticking and bouncing and the much narrower transition width on our understanding of the formation of small bodies in the solar nebular is difficult to predict. Also two potential limitations of our results should be considered.
The experiments were conducted with two different dust analog materials, both consisting of silica but one with a monodisperse and one with a wider size distribution. While the exact material is supposed to have a negligible impact on the experiment outcome (e.g. \citet{BlumWurm:2008}), the size distribution and the monomer size and shape clearly have. It is assumed that the dust population in the protoplanetary disks consists of particles similar to today's interstellar dust grains, with sizes from the nanometer range up to about 1 micrometer. \citep{MathisEtal:1977}. However, the abundance of even larger grains is not well determined. Studies of impacts of interstellar dust grains in the solar system \citep{Landgrafetal:2000}, of primitive, chondritic meteorites \citep{ScottKrot:2005} and the samples returned from comet 81P/Wild 2 by the Stardust mission \citep{BrownleeEtal:2006} showed larger dust grains, up to a size of $\sim 10 \mu$m. Assuming that these particles are the remains of the initial protoplanetary dust population, we can assume grain sizes not entirely different from the ones used in our experiment. An estimation of the influence of the monomer size on the sticking velocity is difficult. In principle, a larger diameter corresponding to the average monomer mass of the polydisperse analog dust (2.0 $\mu$m) should lead to a lower maximum sticking velocity compared to the monodisperse dust (1.5 $\mu$m) (compare e.g. the sticking / fragmentation transition in \citet{DominikTielens:1997}). However, due to the limited data it is not possible to see a clear trend in our data. Figure \ref{P_Transition} shows that the agglomerates made out of polydisperse dust (red diamonds) stick at slightly lower velocities than the agglomerates made of monodisperse particles (green dots), which is in agreement with theory. However, a model-based-, quantitative comparison of the two threshold velocities is, unfortunately, not feasible, as theoretical models assume spherical dust grains. This is not the case for our polydisperse dust samples which consist of irregular shaped grains.\\
A second point which should be addressed is the single porosity of our samples ($\phi \sim 0.37$). Various numerical simulations \citep{SuyamaEtal:2008, ZsomEtal:2011, OkuzumiEtal:2012}  have shown that the first dust agglomerates might have had a much lower volume filling factor (down to $\phi\sim 10^{-3}$). \citet{ZsomEtal:2011} showed that particles with a mass of $\sim 10^{-6}$ g can still have such a low density. Their simulations also showed that, after a relatively short time (few hundred years), these particles can become more compact and end with values comparable to our samples. However, this is based on the compaction model from \citet{GuettlerEtal:2010} and \citet{WeidlingEtal:2009} who already assume spherical compacted ( $\phi \gtrsim 0.1$) particles. Our experiments showed that even aggregates of a few hundred micrometers in size are more likely to form open structures than spherical objects and thus have a lower filling factor than previously assumed. The formation of very small, fractal agglomerates had also been observed in experiments studying an earlier growth regime \citep{WurmBlum:1998, BlumEtal:2002}. In our case, collisions of aggregate clusters, which led to sticking in most cases, caused hardly any visible compaction. This fact emphasizes that also small agglomerates might have had a very low filling factor for a longer time than predicted in simulations, because they were not in a regime in which bouncing would compact them. Therefore, it should be considered that our new collision model might only be applicable to later phases of planetesimal formation after the dust aggregates got compacted.\\
Considering these constraints, our results suggest that there is a region of increased collisional growth between equal-sized dust aggregates which benefits the formation of larger objects. When dust aggregates possess a collision-velocity distribution rather than a fixed constant value, a combination with the probability distribution of sticking and bouncing can lead to a continuous growth of "lucky winners", as recently shown by \citet{WindmarkEtal:2012b}. Dust-aggregate growth can also benefit from the open structure of the growing dust agglomerates. Current models make the assumption that a dust aggregate of a given mass has a spherical shape and homogenous structure. An open-structured cluster of dust aggregates of the same mass would more likely stick in a collision due to its ability to dissipate more energy in restructuring, as shown in Section \ref{subsection_sticking_of_fracatls}. Such clusters also possess a larger cross section, which makes collisions more likely. This effect might be of interest for future simulations. However, more studies of the fractal dimension and the restructuring of these clusters of dust agglomerates are needed. Fig. \ref{Fig.newmodel} also indicates that there might be a direct transition from sticking to fragmentation for dust aggregates with masses below $\sim 10^{-7}~\rm g$. This mass corresponds to a dust agglomerate which consists of $\sim 10^5$ monomer grains. Such a dust agglomerate is about ten times as massive as the ones simulated by \citet{WadaEtal:2011}, which can possibly explain why they did not observe bouncing for porous dust agglomerates. Anyhow, it is questionable if the onset of fragmentation is totally mass independent. Recent experiments by \citet{BeitzMeisneretal:2011} and \citet{SchraeplerEtal:2012} have shown that dust aggregates larger than one centimeter fragment at velocities slightly lower than 1 \metersecond. This, however, does not change the picture qualitatively.

\section{Conclusion}\label{section_conclusion}
We measured the outcome of collisions between sub-mm-sized dust agglomerates at low velocities of several $\mathrm{cm~s^{-1}}$ under micro-gravity conditions. We were able to confirm the findings of earlier experiments with larger dust aggregates that a transition region between perfect sticking and perfect bouncing exists. Based on our experiments and those from Paper I, we determined a new power law for the mass-velocity dependence of the 50\%/50\% sticking/bouncing value as well as the width of the transition from 100\%/0\% to 0\%/100\% sticking/bouncing. This power law is shallower than assumed in Paper I which is based on the model by \citet{GuettlerEtal:2010}. This suggests that growth due to sticking is possible even at velocities higher than before assumed. This disagreement also shows that the collision physics between porous agglomerates can not be easily described by modified contact models for solid spheres.
We also found that clusters of dust aggregates even stick at each other at velocities where compact dust agglomerates would only bounce. Previous collision models, which were used in growth simulations, treated the newly formed agglomerates as compact spheres. Considering our observation that open structures benefit the growth significantly might help to overcome the various barriers found in simulations.
These clusters can either be rather spherical or open-structured. For the latter species, we analyzed the morphology of one example and found that its fractal dimension is on the order of $\mathrm{ D_f = 1.7}$. Furthermore, we analyzed the structure of one of our dust aggregates using nano-CT imaging. We confirmed previous measurements of the filling factor to be about $\phi = 0.37$ and were not able to find any compacted rim that had been suggested to be a reason for the bouncing of dust aggregates. We also estimated the coordination number of these dust aggregates. Using different relations from the literature, it is most likely that our dust aggregates possess a coordination number well below $\mathrm{n_c}=6$, which is assumed to be the critical value at which simulations showed bouncing. Our new data for the sticking-bouncing threshold suggest that a direct transition from sticking to fragmentation, as found in numerical simulations, is possible for dust aggregates consisting of less than $\sim 10^5$ monomer grains, which is the range of the numerical simulations. Thus, the non-occurrence of bouncing could be a finite-size effect.

 \textit{Acknowledgments} We thank the Deutsche Forschungsgemeinschaft for funding this work within the Forschergruppe 759 ``The Formation of Planets: The Critical First Growth Phase'' under grant Bl 298/14-1 and the Deutsches Zentrum für Luft- und Raumfahrt for providing us with drop-tower flights. We also like to thank Olga Mursajew for many interesting discussion about the use of finance mathematics in physics. The nano-CT images were taken at RJL Micro \& Analytic GmbH. We thank Markus J. Heneka and the staff of RJL Micro \& Analytic GmbH for their support. We also like to thank Stephan Olliges from the Institute for Particle Technology for measuring the size distribution of our samples.

\bibliographystyle{aa}
\bibliography{literatur}

\end{document}